\documentclass[english,aps,preprint]{revtex4}
\usepackage[T1]{fontenc}
\usepackage[latin9]{inputenc}
\usepackage{color}
\usepackage{mathrsfs}
\usepackage{amsmath}
\usepackage{amssymb}
\usepackage{graphicx}

\makeatletter
\@ifundefined{textcolor}{}
{%
 \definecolor{BLACK}{gray}{0}
 \definecolor{WHITE}{gray}{1}
 \definecolor{RED}{rgb}{1,0,0}
 \definecolor{GREEN}{rgb}{0,1,0}
 \definecolor{BLUE}{rgb}{0,0,1}
 \definecolor{CYAN}{cmyk}{1,0,0,0}
 \definecolor{MAGENTA}{cmyk}{0,1,0,0}
 \definecolor{YELLOW}{cmyk}{0,0,1,0}
}

\makeatother

\usepackage{babel}
\begin{document}
\title{Nonequilibrium free energy and information flow of a double quantum-dot
system with Coulomb coupling}
\author{Zhiyuan Lin}
\address{Department of Physics, Xiamen University, Xiamen 361005, People's
Republic of China}
\author{Tong Fu}
\address{Department of Physics, Xiamen University, Xiamen 361005, People's
Republic of China}
\author{Juying Xiao}
\address{Department of Physics, Xiamen University, Xiamen 361005, People's
Republic of China}
\author{Shanhe Su}
\email{shanhesu@xmu.edu.cn}

\address{Department of Physics, Xiamen University, Xiamen 361005, People's
Republic of China}
\author{Jincan Chen}
\address{Department of Physics, Xiamen University, Xiamen 361005, People's
Republic of China}
\author{Yanchao Zhang}
\email{zhangyanchao@gxust.edu.cn}

\address{School of Science, Guangxi University of Science and Technology, Liuzhou
545006, People's Republic of China}
\begin{abstract}
We build a double quantum-dot system with Coulomb coupling and aim
at studying the connections among the entropy production, free energy,
and information flow. By utilizing the concepts in stochastic thermodynamics
and graph theory analysis, the Clausius and nonequilibrium free energy
inequalities are built to interpret the local second law of thermodynamics
for subsystems. A fundamental set of cycle fluxes and affinities is
identified to decompose the two inequalities by using Schnakenberg's
network theory. The results show that the thermodynamic irreversibility
has the energy-related and information-related contributions. A global
cycle associated with the feedback-induced information flow would
pump electrons against the bias \textcolor{black}{voltage, which implements
a Maxwell Demon.}

\end{abstract}
\maketitle

\section{INTRODUCTION}

Irreversible thermodynamics is an extension of thermodynamics that
studies the transport phenomena such as the exchanges of mass, energy,
and charge \cite{RN21}. The Onsager reciprocity theorem expresses
the rate of entropy production as the sum of the products of each
flux and its conjugate affinity \cite{RN19,RN20}. This thermodynamic
relation is applicable for most far-from-equilibrium systems, even
without accounting for the linear response regime. A nonzero affinity
implies that a system is not in equilibrium and irreversible processes
drive the system towards the state of equilibrium. The criterion for
the selection of the basic thermodynamic flows and forces remains
a question worth exploring. For instance, the rate of entropy production
of a thermoelectric device can be equivalently expressed in energy
and heat representations \cite{RN22}. Instead of collecting flows
according to the thermodynamic forces acting in the energy transduction,
biogeochemical systems adopt the basic graph concepts used in network
analysis \cite{RN23}.

Graph theory analysis indicates that each cycle makes an additive
positive contribution to the total rate of entropy production in the
ensemble \cite{RN24}. Schnakenberg expressed the macroscopic entropy
production of stochastic processes in terms of the cycles from the
network, because the products of the transition rates along a cycle
depend only on the macroscopic thermodynamic affinities maintaining
the system out of equilibrium \cite{RN25}. Horowitz et al. used a
graph theoretic method to provide a unified thermodynamic scheme describing
information transfers in autonomous systems \cite{RN11}. Yamamoto
introduced a graph contraction method to prove that the Onsager coefficient
associated with the driving of an information current satisfies the
Onsager reciprocity \cite{RN10}. Graph theory concepts have achieved
great success at learning irreversible thermodynamics of the energy
\cite{RN28,RN29}, entorpy, fluctuation \cite{RN9}, and information
at nanoscale\cite{RN46,RN47}.

On the one hand, many studies were concerned with the formalism of
free energy for irreversible thermodynamics. Crooks related nonequilibrium
measurements of free energy differences to the work done on microscopically
reversible Markovian systems \cite{RN31,RN32}. The Jarzynski relation,
which relates the free energy differences between two sta\textcolor{black}{tes
to the irr}eversible work along an ensemble of trajectories joining
the same states, has been often used for calculating the equilibrium
free energies of classical and quantum systems \cite{RN13,RN34,RN35,RN36}.
Esposito introduced the concept of nonequilibrium system free energy
to understand the irreversible work in Hamiltonian dynamics of an
open driven system \cite{RN17,RN38}. More recently, the investigations
of free energy were generalized to the systems coupled to the environment
with multiple heat baths \cite{RN18,RN49,RN50,RN39}. More and More
studies utilized the nonequilibrium Clausius and free energy inequalities
to clarify the information and energy regimes in the nonequilibrium
systems \cite{RN51,RN52,RN53,RN54,RN55,RN56,RN57,RN58,RN59}. Miyahara
et al. derived the Sagawa-Ueda-Jarzynski relation under a nonisothermal
system to measure the change of the free energy \cite{RN39}. Ptaszy\'{n}ski
et al. formulated a nonequilibrium free energy inequality for a generic
open quantum system weakly coupled to multi heat sources \cite{RN18}.
Despite recent developments, relatively little attention has been
paid to correlate the non-equilibrium free energy in terms of thermodynamic
affinities and flows. By considering the great success of graph theory
analysis in irreversible thermodynamics, a cycle decomposition may
help to establish this connection.

Considering a double-quantum-dot system, we will employ graph theory
to analyze the entropy production and the non-equilibrium free energy
of the open quantum system obeying continuous-time Markov jump process.
The contents are organized as follows: In Section II, the general
model of two quantum dots coupled in parallel to four electronic reservoirs
with different chemical potentials is briefly described. Schnakenberg's
network theory are applied to obtain the fundamental cycle fluxes
and affinities. The nonequilibrium Clausius and free energy inequalities
of the subsystems at steady state are derived. In Section III, the
main thermodynamic characteristic functions as functions of the Coulomb
coupling strength will be evaluated numerically. Finally, the main
conclusions are drawn.
\begin{center}
\begin{figure}
\includegraphics[scale=0.25]{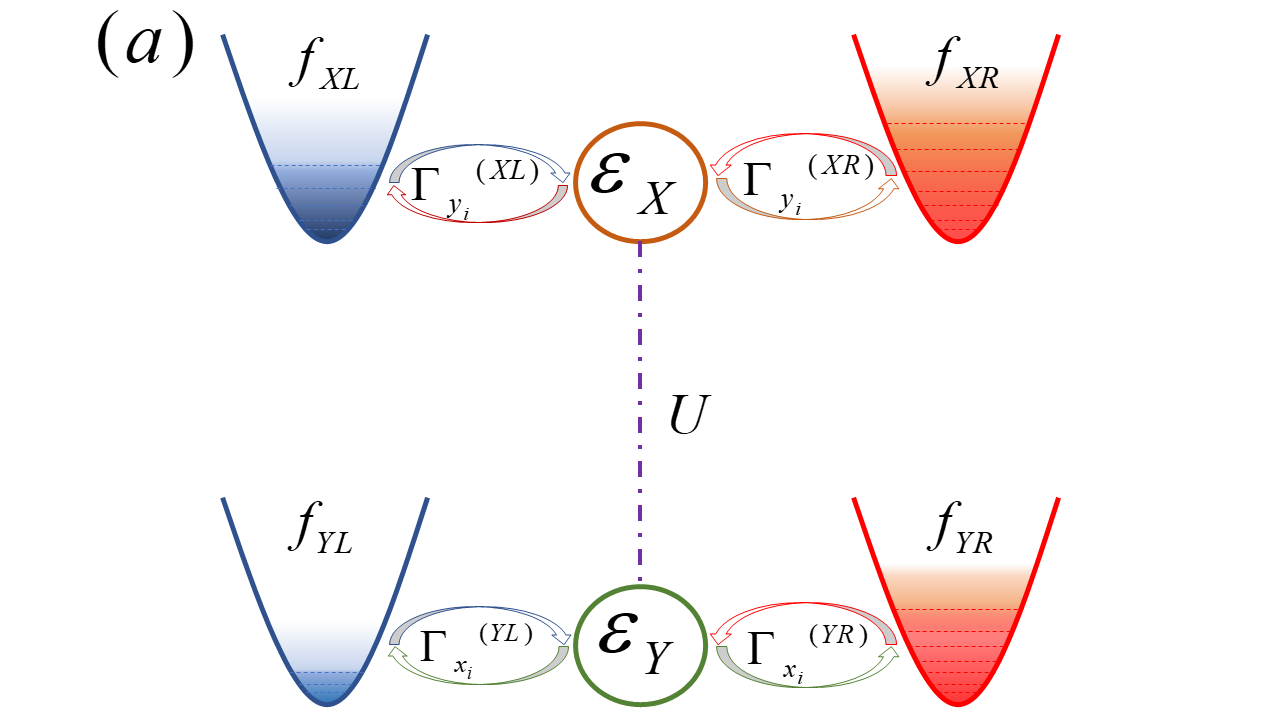}\includegraphics[scale=0.25]{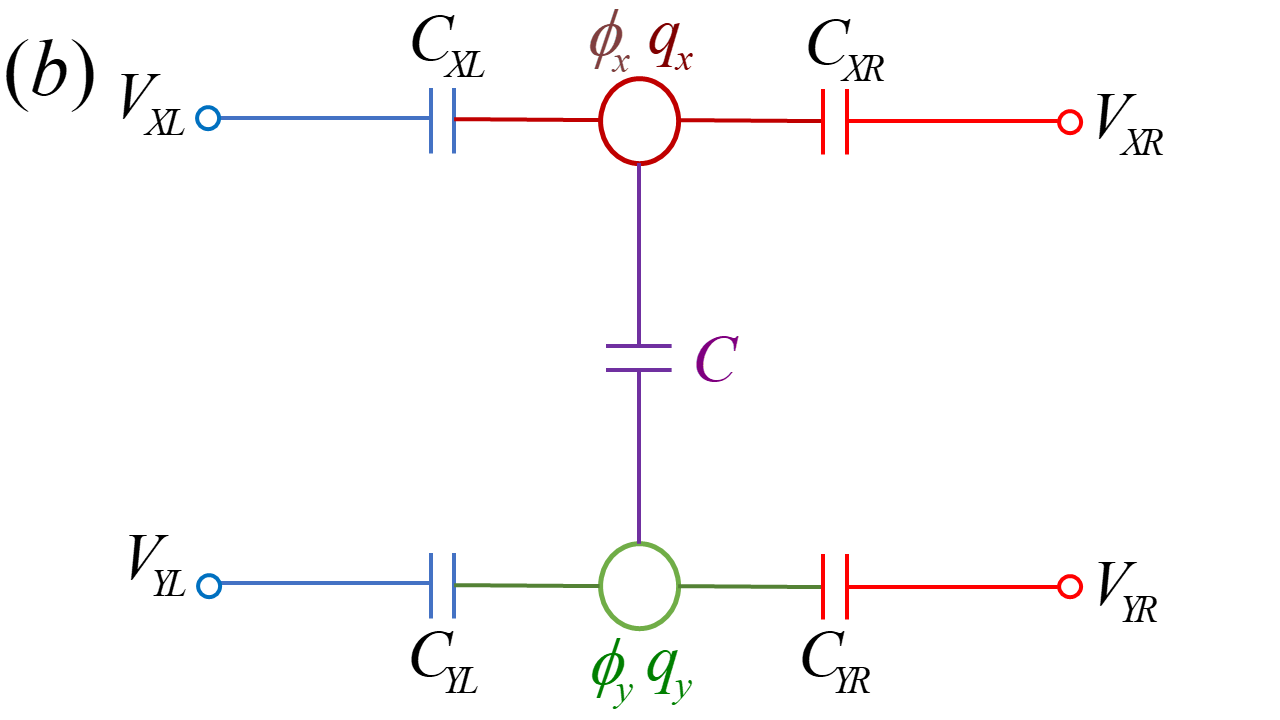}\caption{(a) The Schematic diagram of a Coulomb-coupled double quantum-dot
system. (b) The equivalent circuit showing the capacitive couplings
between the quantum dots and the reservoirs.}
\end{figure}
\par\end{center}

\section{MODEL AND METHODS}

\subsection{\textit{The double quantum-dot system with Coulomb coupling}}

The considered model is a bipartite system that consists of capacitively
coupled quantum dots $X$ and $Y$, as shown in Fig. 1 (a). The quantum
dots $X$ and $Y$ with energies $\varepsilon_{X}$ and $\varepsilon_{Y}$
are coupled via the long-range Coulomb force $U$ such that they only
exchange energy but no electrons. The states of the composite system
are given by a four-state basis $z=\left\{ \left|(x_{0},y_{0})\right\rangle ,\left|\left(x_{1},y_{0}\right)\right\rangle ,\left|\left(x_{0},y_{1}\right)\right\rangle ,\left|\left(x_{1},y_{1}\right)\right\rangle \right\} $,
where $x_{0}$ and $y_{0}$ ($x_{1}$ and $y_{1}$) represent that
the site of $X$ and $Y$ is empty (filled), respectively. The quantum
dot $X$ ($Y$) is connected to two Fermi reservoirs $XL$ and $XR$
($YL$ and $YR$) and admits the electron transports through parallel
interacting channels. Figure 1(b) is the equivalent circuit showing
the capacitive couplings between the quantum dots and the reservoirs.
$V_{\nu}=\mu_{\nu}/e$ ($\nu=XL,XR,YL,YR$) determines the voltage
of the reservoir $\nu$, where $\mu_{\nu}$ is the chemical potential
and $e$ is elementary positive charge. The imbalance capacitances
between the quantum dot and the terminal $\nu$ and between the quantum
dots are defined as $C_{\nu}$ and $C$, respectively. $\phi_{x}$
and $\phi_{y}$ are the electrostatic potentials in each quantum dot
at state $\left|\left(x,y\right)\right\rangle $ . The total charges
of $X$ and $Y$ at state $\left|\left(x,y\right)\right\rangle $
are the sum of the charges on all of the capacitors connected to $X$
and $Y$ \cite{RN63,RN64}, i.e.,
\begin{align}
q_{x} & =C_{XL}\left(\phi_{x}-V_{XL}\right)+C_{XR}\left(\phi_{x}-V_{XR}\right)+C\left(\phi_{x}-\phi_{y}\right),\\
q_{y} & =C_{YL}\left(\phi_{y}-V_{YL}\right)+C_{YR}\left(\phi_{y}-V_{YR}\right)+C\left(\phi_{y}-\phi_{x}\right).
\end{align}

\noindent These two equations can be expressed more compactly in a
matrix form

\begin{equation}
\left[\begin{array}{c}
q_{x}+C_{XL}V_{XL}+C_{XR}V_{XR}\\
q_{y}+C_{YL}V_{YL}+C_{YR}V_{YR}
\end{array}\right]=\left[\begin{array}{cc}
C_{X}^{'} & -C\\
-C & C_{Y}^{'}
\end{array}\right]\left[\begin{array}{c}
\phi_{x}\\
\phi_{y}
\end{array}\right],
\end{equation}

\noindent where $C_{X}^{'}=C_{XL}+C_{XR}+C$ and $C_{Y}^{'}=C_{YL}+C_{YR}+C$
define the total capacitances of $X$ and $Y$. The electrostatic
potentials\textcolor{red}{{} }$\phi_{x}$ and $\phi_{y}$ are then conveniently
expressed by using the capacitance matrix

\begin{equation}
\left[\begin{array}{c}
\phi_{x}\\
\phi_{y}
\end{array}\right]=\frac{1}{C_{X}^{'}C_{Y}^{'}-C^{2}}\left[\begin{array}{cc}
C_{Y}^{'} & C\\
C & C_{X}^{'}
\end{array}\right]\left[\begin{array}{c}
q_{x}+C_{XL}V_{XL}+C_{XR}V_{XR}\\
q_{y}+C_{YL}V_{YL}+C_{YR}V_{YR}
\end{array}\right].
\end{equation}

\textcolor{black}{The electrostatic }energy\textcolor{black}{{} for
a given quantum state is computed by}

\textcolor{black}{
\begin{equation}
U(x,y)=\frac{1}{2}\left[q_{x}+C_{XL}V_{XL}+C_{XR}V_{XR}\text{, \ensuremath{q_{y}}+\ensuremath{C_{YL}V_{YL}}+\ensuremath{C_{YR}V_{YR}}}\right]\left[\begin{array}{c}
\phi_{x}\\
\phi_{y}
\end{array}\right].
\end{equation}
}

\noindent \textcolor{black}{Note that $q_{x}\left(q_{y}\right)=0$
or }$e$\textcolor{black}{, depending on whether $X\left(Y\right)$
is empty} or occupied. Thus, the electrostatic energies for the four
quantum states read
\noindent \begin{flushleft}
\begin{align}
U(x_{0},y_{0}) & =\frac{1}{2\left(C_{X}^{'}C_{Y}^{'}-C^{2}\right)}\left(C_{Y}^{'}\mathcal{Q}_{x}^{2}+2C\mathcal{Q}_{x}\mathcal{Q}_{y}+C_{X}^{'}\mathcal{Q}_{y}^{2}\right),\nonumber \\
U\left(x_{1},y_{0}\right) & =\frac{1}{2\left(C_{X}^{'}C_{Y}^{'}-C^{2}\right)}\left[C_{Y}^{'}\left(e^{2}+2e\mathcal{Q}_{x}\right)+2eC\mathcal{Q}_{y}\right]+U_{x_{0}y_{0}},\nonumber \\
U\left(x_{0},y_{1}\right) & =\frac{1}{2\left(C_{X}^{'}C_{Y}^{'}-C^{2}\right)}\left[2eC\mathcal{Q}_{x}+C_{X}^{'}\left(e^{2}+2e\mathcal{Q}_{y}\right)\right]+U_{x_{0}y_{0}},\\
U\left(x_{1},y_{1}\right) & =\frac{2e^{2}C}{2\left(C_{X}^{'}C_{Y}^{'}-C^{2}\right)}+U_{x_{1}y_{0}}+U_{x_{0}y_{1}}-U_{x_{0}y_{0}},\nonumber
\end{align}
\par\end{flushleft}

\noindent where $\mathcal{Q}_{x}=C_{XL}V_{XL}+C_{XR}V_{XR}$ and $\mathcal{Q}_{y}=C_{YL}V_{YL}+C_{YR}V_{YR}$.
We are now capable of determining the change of energy in the system
when an electron tunnels into a quantum dot. When the other dot is
empty, the charging energies of $X$ and $Y$ are, respectively, given
by

\begin{equation}
U_{X_{0}}=U\left(x_{1},y_{0}\right)-U(x_{0},y_{0}),U_{Y_{0}}=U\left(x_{0},y_{1}\right)-U(x_{0},y_{0}).
\end{equation}
On the other hand, when the other dot is occupied, the charging energies
of $X$ and $Y$ are, respectively, described by
\begin{equation}
U_{X_{1}}=U\left(x_{1},y_{1}\right)-U\left(x_{0},y_{1}\right),U_{Y_{1}}=U\left(x_{1},y_{1}\right)-U\left(x_{1},y_{0}\right).
\end{equation}

\noindent The differences of the charging energies $U_{X_{1}}-U_{X_{0}}=U_{Y_{1}}-U_{Y_{0}}=U$,
where $U=\frac{q^{2}C}{\left(C_{X}^{'}C_{Y}^{'}-C^{2}\right)}$ determines
the quantized energy which can be transferred from one dot to the
other dot.

\subsection{\textit{The master equation}}

Let $p(z,t)$ be the probability of state $z$ of the coupled quantum
dots at time $t$. In the regime of sequential tunneling approximation,
the broadening of energy levels can be neglected and the transmission
through tunnel barriers is defined by the sequential tunneling of
a single electron. Thus, the time evolution of $p(z,t)$ is governed
by a Markovian master equation \cite{RN61,RN62,RN45}

\noindent
\begin{equation}
\frac{d}{dt}p(z,t)=\sum_{z',\nu}\left[R_{zz'}^{(\nu)}p(z',t)-R_{z'z}^{(\nu)}p(z,t)\right].\label{eq:me}
\end{equation}
The transition rate from state $z'$ to state $z$ induced by the
reservoir $\nu$ reads

\begin{equation}
R_{zz^{\prime}}^{(\nu)}=\left\{ \begin{array}{ll}
{R_{xx^{\prime}|y}^{(\nu)}} & {\left(x\neq x^{\prime},y=y^{\prime},\nu=XL,XR\right)}\\
{R_{yy^{\prime}|x}^{(\nu)}} & {\left(x=x^{\prime},y\neq y^{\prime},\nu=YL,YR\right)}\\
{0} & {\text{(otherwise) }}
\end{array}\right.,\label{eq:tr}
\end{equation}

\noindent where $x^{\prime}\in\left\{ x_{0},x_{1}\right\} $ and $y^{\prime}\in\left\{ y_{0},y_{1}\right\} $.
We have assumed that the two subsystems should not change their states
simultaneously during a single transition process. The specific forms
of the transition rates are

\begin{equation}
\begin{array}{l}
{R_{x_{1}x_{0}|y_{i}}^{(\nu)}=\Gamma_{y_{i}}^{(\nu)}f_{y_{i}}^{(\nu)}},\\
{R_{x_{0}x_{1}|y_{i}}^{(\nu)}=\Gamma_{y_{i}}^{(\nu)}\left(1-f_{y_{i}}^{(\nu)}\right)},\\
{R_{y_{1}y_{0}|x_{i}}^{(\nu)}=\Gamma_{x_{i}}^{(\nu)}f_{x_{i}}^{(\nu)}},\\
{R_{y_{0}y_{1}|x_{i}}^{(\nu)}=\Gamma_{x_{i}}^{(\nu)}\left(1-f_{x_{i}}^{(\nu)}\right)},
\end{array}\label{eq:TR}
\end{equation}

\noindent where $f_{x_{i}}^{(\nu)}=\left\{ 1+\textrm{exp}\left[\beta_{\nu}\left(\varepsilon'_{Y}+iU-\mu_{\nu}\right)\right]\right\} ^{-1}$,
$f_{y_{i}}^{(\nu)}=\left\{ 1+\textrm{exp}\left[\beta_{\nu}\left(\varepsilon'_{X}+iU-\mu_{\nu}\right)\right]\right\} ^{-1}$$(i=0,1)$,
$\varepsilon'_{X}=\varepsilon_{X}+U_{X_{0}}$, and $\varepsilon'_{Y}=\varepsilon_{Y}+U_{Y_{0}}$.
The reservoir $\nu$ is at temperature $T_{\nu}$ and the chemical
potential is $\mu_{\nu}$. All temperatures are set to be equal, i.e.,
$T_{\nu}=T$. The inverse temperature parameter $\beta_{\nu}=1/\left(k_{B}T_{\nu}\right)$,
where $k_{B}$ is Boltzmann's constant and we set $k_{B}=1$ in the
discussion. $\Gamma_{x_{i}}^{(\nu)}$ ($\Gamma_{y_{i}}^{(\nu)})$
is a positive constant describing the height of the potential barrier
between the dot $Y(X)$ and the reservoir $\nu$. The potential barrier
of $Y$, characterized by $\Gamma_{x_{i}}^{(\nu)}$, depends on the
state of $X$, and vice versa.
\begin{center}
\begin{figure}
\includegraphics[scale=0.3]{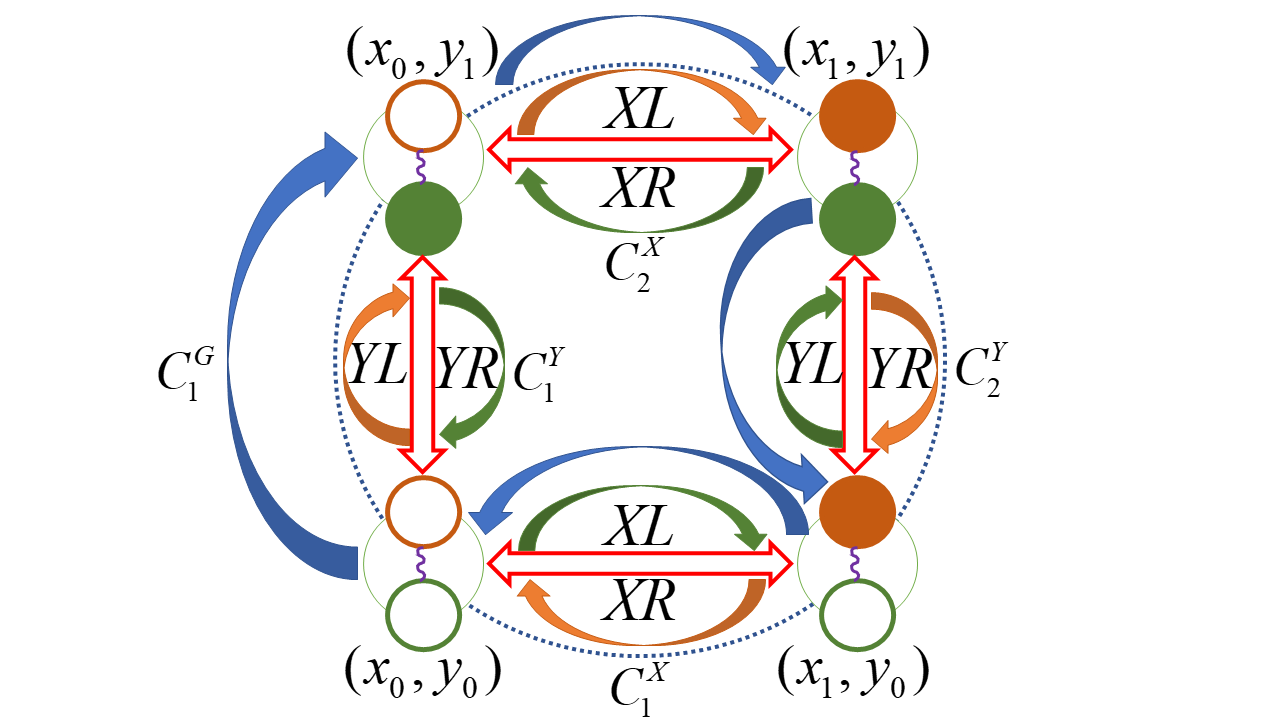}\caption{Illustration of the directed graph and the cycle basis. The vertices
represent the four states of the two quantum dots, and the double-headed
arrows stand for the forward and backward transitions between the
two states induced by reservoirs. The single-headed curly arrows are
the edges of a cycle basis with a directed orientation. Note that
we can arbitrarily choose the direction of an edge. \label{fig:graph}}
\end{figure}
\par\end{center}

\subsection{\textit{Schnakenberg's network theory}}

Schnakenberg stated that nonequilibrium random processes could be
investigated and understood by carrying out the graph analysis associated
with the master equation \cite{RN8,RN9,RN10}. For purposes of relating
the thermodynamic properties to the fundamental fluxes and affinities,
a graph representation of the dynamics of the double quantum-dot system
is introduced (Fig. 2).

According to Schnakenberg's network theory, a vertex in the graph
represents a state of the system $z$. A directed edge $e:=\left(z'\stackrel{\nu}{\rightarrow}z\right)$
is corresponding to the transition from state $z'$ to state $z$
through the reservoir $\nu$ with a nonzero transition rate $R_{zz'}^{(v)}$.
For edge $e$, the edge affinity and edge current are identified as

\begin{equation}
\mathcal{F}_{e}=\ln\frac{R_{zz^{\prime}}^{(\nu)}}{R_{z^{\prime}z}^{(\nu)}},\label{eq:affi}
\end{equation}

\begin{equation}
J_{e}=R_{zz'}^{(\nu)}p(z',t)-R_{z'z}^{(\nu)}p(z,t),\label{eq:ec}
\end{equation}
respectively. We next define the effective affinity of edge $e$

\begin{equation}
\mathcal{\mathcal{\mathscr{F}}}_{e}:=\mathcal{F}_{e}+\ln\frac{p(z',t)}{p(z,t)}\text{.}\label{eq:effaff}
\end{equation}
Because $J_{e}=0$ resulting in $\mathcal{\mathscr{F}}_{e}=0$, the
effective affinity acts as the conjugate of the current $J_{e}$.
In contrast, $\mathcal{F}_{e}$ can be regarded as the bare thermodynamic
affinity related to the detailed balance.

According to Kirchhoff's current law \cite{RN8}, the stationary Master
equation {[}Eq. (\ref{eq:me}) with $\frac{d}{dt}p(z,t)=0${]} leads
to

\begin{equation}
\sum_{z',\nu}J_{e}=0.\label{eq:me-1}
\end{equation}
By specifying the set $\mathcal{E}$ as a collection of the directed
edges (represented by the single-headed curly arrows) in Fig. 2, we
have $\left(z^{\prime}\rightarrow z\right)\in\mathcal{E}\Rightarrow\left(z\rightarrow z^{\prime}\right)\notin\mathcal{E}$.
Under the assignment in Eq. (\ref{eq:tr}), each edge only describes
a transition in $X$ or $Y$. Therefore, one can use the set $\mathcal{E}^{X}:=\left\{ e=\left(\left(x^{\prime},y\right)\stackrel{v}{\rightarrow}(x,y)\right)\in\mathcal{E}\right\} \left(x\neq x^{\prime}\right)$
to describe the transitions in $X$ and the other set $\mathcal{E}^{Y}:=\left\{ e=\left(\left(x,y^{\prime}\right)\stackrel{\nu}{\rightarrow}(x,y)\right)\in\mathcal{E}\right\} \left(y\neq y^{\prime}\right)$
to cover the transitions in $Y$. The set $\mathcal{E}$ of all edges
is then divided into two parts $\mathcal{E}^{X}$ and $\mathcal{E}^{Y}$.

The cycle basis of the directed graph (Fig. 2) $\mathcal{C}=\left\{ C_{1}^{X},C_{2}^{X},C_{1}^{G},C_{1}^{Y},C_{2}^{Y}\right\} $
may be classified into the local and global cycles as

\begin{align}
C_{1}^{X} & :=\left\{ \left(x_{0},y_{0}\right)\stackrel{XL}{\rightarrow}\left(x_{1},y_{0}\right)\stackrel{XR}{\rightarrow}\left(x_{0},y_{0}\right)\right\} ,\nonumber \\
C_{2}^{X} & :=\left\{ \left(x_{0},y_{1}\right)\stackrel{XL}{\rightarrow}\left(x_{1},y_{1}\right)\stackrel{XR}{\rightarrow}\left(x_{0},y_{1}\right)\right\} ,\nonumber \\
C_{1}^{Y} & :=\left\{ \left(x_{0},y_{0}\right)\stackrel{YL}{\rightarrow}\left(x_{0},y_{1}\right)\stackrel{YR}{\rightarrow}\left(x_{0},y_{0}\right)\right\} ,\label{eq:clc}\\
C_{2}^{Y} & :=\left\{ \left(x_{1},y_{0}\right)\stackrel{YL}{\rightarrow}\left(x_{1},y_{1}\right)\stackrel{YR}{\rightarrow}\left(x_{1},y_{0}\right)\right\} ,\nonumber \\
C_{1}^{G} & :=\left\{ \left(x_{0},y_{0}\right)\stackrel{YL}{\rightarrow}\left(x_{0},y_{1}\right)\stackrel{XL}{\rightarrow}\left(x_{1},y_{1}\right)\stackrel{YL}{\rightarrow}\left(x_{1},y_{0}\right)\stackrel{XL}{\rightarrow}\left(x_{0},y_{0}\right)\right\} .\nonumber
\end{align}

\noindent A directed cycle is a directed sequence of the connected
edges with the same initial and terminal vertexes. This set of cycles\textcolor{black}{{}
is} broadly classified into three groups: local cycles of $X$, $\mathcal{C}^{X}:=\left\{ C_{1}^{X},C_{2}^{X}\right\} $;
local cycles of $Y$, $\mathcal{C}^{Y}:=\left\{ C_{1}^{Y},C_{2}^{Y}\right\} $;
and a global cycle, $\mathcal{C}^{G}:=C_{1}^{G}$ . A local cycle
indicates that $X\left(Y\right)$ is fixed and $Y\left(X\right)$
is changing, which supports the internal subsystem flows. The global
cycle $C_{1}^{G}$ links $X$ and $Y$, so that a current flowing
around a global cycle carries energy and entropy from one subsystem
to the other. Any other cycles are recognized as a linear combination
of cycles in $\mathcal{C}$ \cite{RN11}. It will be useful to introduce
the function

\begin{equation}
\Xi\left(e,C_{k}\right):=\left\{ \begin{array}{ll}
1 & \left(e\in C_{k}^{l}\right)\\
-1 & \left(e^{\dagger}\in C_{k}^{l}\right)\\
0 & \left(\mathrm{otherwise}\right)
\end{array},\right.\label{eq:mt}
\end{equation}
where $e\in C_{k}^{l}\left(l=X,Y,G;k=1,2\right)$ means that $e$
is one of the edges in $C_{k}^{l}$, and $e^{\dagger}:=\left(z\stackrel{\nu}{\rightarrow}z^{\prime}\right)$
defines the backward transition edge of $e$. To each fundamental
cycle, one can then assign the affinity $\mathcal{F}\left(C_{k}^{l}\right)$
as a sum of the affinities along the edges in $C_{k}^{l}$

\begin{equation}
\mathcal{F}\left(C_{k}^{l}\right):=\sum_{e\in\mathcal{E}}\Xi\left(e,C_{k}^{l}\right)\mathcal{F}_{e}\text{.}\label{eq:caf}
\end{equation}
The motive is to introduce the partial affinities of the cycles associated
with $X$ and $Y$ by

\begin{equation}
\mathcal{\mathscr{\mathcal{F}}}_{X}\left(C_{k}^{l}\right)=\sum_{e\in\mathcal{E}^{X}}\Xi\left(e,C_{k}^{l}\right)\mathcal{\mathcal{\mathscr{\mathcal{F}}}}_{e},\label{eq:cal-2-1}
\end{equation}

\begin{equation}
\mathcal{\mathscr{\mathcal{F}}}_{Y}\left(C_{k}^{l}\right)=\sum_{e\in\mathcal{E}^{Y}}\Xi\left(e,C_{k}^{l}\right)\mathcal{\mathcal{\mathscr{\mathcal{F}}}}_{e}.\label{eq:cal-3-1}
\end{equation}
Note that the partial affinities are found to be as dissimilar as
$\mathcal{F}\left(C_{k}^{l}\right)$, because the sums are not taken
over $\mathcal{E}$ but over $\mathcal{E}^{X}$ and $\mathcal{E}^{Y}$.

By using the definition in Eq. (\ref{eq:effaff}), one can also define
the effective affinity of $C_{k}^{l}$ by

\begin{equation}
\mathscr{F}\left(C_{k}^{l}\right):=\sum_{e\in\mathcal{E}}\Xi\left(e,C_{k}^{l}\right)\mathcal{\mathcal{\mathscr{F}}}_{e}\text{.}\label{eq:caf-1}
\end{equation}
The partial effective affinities of the cycles corresponding to $X$
and $Y$ are written as

\begin{equation}
\mathcal{\mathscr{F}}_{X}\left(C_{k}^{l}\right)=\sum_{e\in\mathcal{E}^{X}}\Xi\left(e,C_{k}^{l}\right)\mathcal{\mathcal{\mathscr{F}}}_{e},\label{eq:cal-2}
\end{equation}

\begin{equation}
\mathcal{\mathscr{F}}_{Y}\left(C_{k}^{l}\right)=\sum_{e\in\mathcal{E}^{Y}}\Xi\left(e,C_{k}^{l}\right)\mathcal{\mathcal{\mathscr{F}}}_{e}.\label{eq:cal-3}
\end{equation}
Combining Eqs. (\ref{eq:caf-1})-(\ref{eq:cal-3}) with Eqs. (\ref{eq:affi})
and (\ref{eq:mt}), one can obtain the effective affinities of local
cycles and the partial effective affinities of the global cycle in
Eq. (\ref{eq:clc})

\begin{equation}
\mathcal{\mathscr{F}}(C_{1}^{X})=\mathcal{\mathscr{F}}(C_{2}^{X}):=-\Delta\mu_{X}/T,
\end{equation}
\begin{equation}
\mathcal{\mathscr{F}}(C_{1}^{Y})=\mathcal{\mathscr{F}}(C_{2}^{Y}):=-\Delta\mu_{Y}/T,
\end{equation}

\begin{equation}
\mathcal{\mathscr{F}}_{X}\left(C_{1}^{G}\right)=-\mathcal{\mathscr{F}}_{Y}\left(C_{1}^{G}\right)=\ln\frac{p\left(x_{0},y_{1}\right)p\left(x_{1},y_{0}\right)}{p\left(x_{0},y_{0}\right)p\left(x_{1},y_{1}\right)}-U/T\text{,}\label{eq:peag}
\end{equation}
where
$\Delta\mu_{X}=\mu_{XR}-\mu_{XL}$ and $\Delta\mu_{Y}=\mu_{YR}-\mu_{YL}$.
For the global cycle $C_{1}^{G}$, the partial affinities $\mathcal{\mathscr{\mathcal{F}}}_{X}\left(C_{1}^{G}\right)=-\mathcal{\mathscr{\mathcal{F}}}_{Y}\left(C_{1}^{G}\right)=-U/T$.
The partial effective affinities of the global cycle in Eq. (\ref{eq:peag})
is equal to the sum of the two forces, i.e.,
\begin{equation}
\mathcal{\mathscr{F}}_{X}\left(C_{1}^{G}\right)=\mathcal{\mathscr{\mathcal{F}}}_{X}\left(C_{1}^{G}\right)+\mathcal{\mathscr{\mathcal{F}}}_{I}\left(C_{1}^{G}\right),
\end{equation}
\begin{equation}
\mathcal{\mathscr{F}}_{Y}\left(C_{1}^{G}\right)=\mathcal{\mathscr{\mathcal{F}}}_{Y}\left(C_{1}^{G}\right)-\mathcal{\mathscr{\mathcal{F}}}_{I}\left(C_{1}^{G}\right),
\end{equation}
where $\mathcal{\mathscr{\mathcal{F}}}_{I}\left(C_{1}^{G}\right)=\ln\frac{p\left(x_{0},y_{1}\right)p\left(x_{1},y_{0}\right)}{p\left(x_{0},y_{0}\right)p\left(x_{1},y_{1}\right)}$
provides the driving force for the information exchanges between $X$
and $Y$.

The transitions around cycle $C_{k}^{l}$ generate the net current
$J\left(C_{k}^{l}\right)$ (cycle current), which is fundamental in
irreversible thermodynamics. At steady state, Kirchhoff's laws govern
the conservation of energy and charge, indicating that

\begin{equation}
J_{e}=\sum_{C_{k}^{l}\in\mathcal{C}}\Xi\left(e,C_{k}^{l}\right)J\left(C_{k}^{l}\right).
\end{equation}

\noindent Therefore, the currents corresponding to the basic cycles
{[}Eq. (\ref{eq:clc}){]} can be calculated as

\begin{align}
J(C_{1}^{X}) & :=R_{x_{0}x_{1}|y_{0}}^{(XR)}p(x_{1},y_{0})-R_{x_{1}x_{0}|y_{0}}^{(XR)}p(x_{0},y_{0})=J_{x_{0}x_{1}|y_{0}}^{(XR)},\nonumber \\
J(C_{2}^{X}) & :=R_{x_{0}x_{1}|y_{1}}^{(XR)}p(x_{1},y_{1})-R_{x_{1}x_{0}|y_{1}}^{(XR)}p(x_{0},y_{1})=J_{x_{0}x_{1}|y_{1}}^{(XR)},\nonumber \\
J(C_{1}^{G}) & :=\left(R_{y_{1}y_{0}|x_{0}}^{(YL)}+R_{y_{1}y_{0}|x_{0}}^{(YR)}\right)p(x_{0},y_{0})-\left(R_{y_{0}y_{1}|x_{0}}^{(YL)}+R_{y_{0}y_{1}|x_{0}}^{(YR)}\right)p(x_{0},y_{1})=J_{y_{1}y_{0}|x_{0}}^{(YL)}+J_{y_{1}y_{0}|x_{0}}^{(YR)},\label{eq:cc}\\
J(C_{1}^{Y}) & :=R_{y_{0}y_{1}|x_{0}}^{(YR)}p(x_{0},y_{1})-R_{y_{1}y_{0}|x_{0}}^{(YR)}p(x_{0},y_{0})=J_{y_{0}y_{1}|x_{0}}^{(YR)},\nonumber \\
J(C_{2}^{Y}) & :=R_{y_{0}y_{1}|x_{1}}^{(YR)}p(x_{1},y_{1})-R_{y_{1}y_{0}|x_{1}}^{(YR)}p(x_{1},y_{0})=J_{y_{0}y_{1}|x_{1}}^{(YR)},\nonumber
\end{align}
where $J_{xx'|y}^{(\nu)}$ represents the current due to the transition
$(x',y)\stackrel{\nu}{\rightarrow}(x,y)$, and similarly for $J_{yy'|x}^{(\nu)}$.

The net electronic currents from $XR$ and $XL$ and from $YR$ and
$YL$ are, respectively, simplified as

\begin{equation}
J_{X}=-\left(J(C_{1}^{X})+J(C_{2}^{X})\right),\label{eq:cx}
\end{equation}

\begin{equation}
J_{Y}=-\left(J(C_{1}^{Y})+J(C_{2}^{Y})\right).\label{eq:cy}
\end{equation}

\subsection{\textit{Nonequilibrium Clausius inequalities of the two subsystems}}

Given the probability of a microstate, the entropy of the system is
as follows \cite{RN12,RN13,RN14}
\begin{equation}
S\left(t\right)=-\sum_{z}p(z,t)\ln p(z,t).\label{eq:gen}
\end{equation}
For steady states, the time derivative of Eq. (\ref{eq:gen}) reduces
to
\begin{equation}
\dot{S}=\dot{\sigma}+\dot{S}_{\mathrm{r}}=0,
\end{equation}
where

\begin{equation}
\dot{\sigma}=\sum_{\nu}\sum_{z,z^{\prime}}R_{zz^{\prime}}^{(\nu)}p\left(z^{\prime},t\right)\ln\frac{R_{zz'}^{(\nu)}p\left(z^{\prime},t\right)}{R_{z'z}^{(\nu)}p(z,t)}\label{eq:epr}
\end{equation}
is the rate of total entropy production \cite{RN15}, and

\begin{equation}
\dot{S}_{r}=-\sum_{\nu}\sum_{z,z^{\prime}}R_{zz^{\prime}}^{(\nu)}p\left(z^{\prime},t\right)\ln\frac{R_{zz'}^{(\nu)}}{R_{z'z}^{(\nu)}}\label{eq:el}
\end{equation}
describes the the entropy flow from reservoirs.

The Logarithmic sum inequality states that for non-negative $a_{i}$
and $b_{i}$, $\sum_{i=1}^{n}a_{i}\ln\frac{a_{i}}{b_{i}}\geq a\ln\frac{a}{b}$
with $a=\sum_{i}a_{i}$ and $b=\sum_{i}b_{i}$ . Therefore, one can
prove that $\dot{\sigma}\geq0$ and the second law of thermodynamics
holds for the system. In the long time limit, the system reaches an
unique non-equilibrium stationary state and $\dot{S}=0$. The rate
of entropy production must be balanced by the entropy flows through
its terminals, i.e., $\dot{\sigma}=-\dot{S}_{r}$.

The transiton rates {[}Eq. (\ref{eq:TR}){]} between different states
satisfy the local detailed balance, e.g., $\ln\frac{R_{x_{1}x_{0}|y_{0}}^{(XR)}}{R_{x_{0}x_{1}|y_{0}}^{(XR)}}=-\left(\varepsilon'_{X}-\mu_{XR}\right)/T$
and $\ln\frac{R_{x_{1}x_{0}|y_{1}}^{(XR)}}{R_{x_{0}x_{1}|y_{1}}^{(XR)}}=-\left(\varepsilon'_{X}+U-\mu_{XR}\right)/T$.
Thus, the entropy flow is simplified as

\begin{equation}
\dot{S}_{r}=-\left[J_{X}\Delta\mu_{X}+J_{Y}\Delta\mu_{Y}\right]/T.\label{eq:el2}
\end{equation}
For the interaction between $X$ and reservoir \textcolor{black}{$XR$,
$\varepsilon'_{X}-\mu_{XR}$ and $\varepsilon'_{X}+U-\mu_{XR}$ represent
the thermal energies supplied by reservoir $XR$ during a jump in
the subsystem $X$, which enable}s us to identify the heat currents
into $X$ from reservoir $XR$

\begin{align}
\dot{Q}_{XR} & =T\left(J_{x_{0}x_{1}|y_{0}}^{(XR)}\ln\frac{R_{x_{1}x_{0}|y_{0}}^{(XR)}}{R_{x_{0}x_{1}|y_{0}}^{(XR)}}+J_{x_{0}x_{1}|y_{1}}^{(XR)}\ln\frac{R_{x_{1}x_{0}|y_{1}}^{(XR)}}{R_{x_{0}x_{1}|y_{1}}^{(XR)}}\right)\nonumber \\
 & =-J_{x_{0}x_{1}|y_{0}}^{(XR)}\left(\varepsilon'_{X}-\mu_{XR}\right)-J_{x_{0}x_{1}|y_{1}}^{(XR)}\left[\left(\varepsilon'_{X}+U-\mu_{XR}\right)\right]\text{.}\label{eq:qxr}
\end{align}
By following the same path, the heat currents into the system from
reservoirs $XL$, $YR$, and $YL$ are, respectively, given by

\begin{equation}
\dot{Q}_{XL}=J_{x_{1}x_{0}|y_{0}}^{(XL)}\left(\varepsilon'_{X}-\mu_{XL}\right)+J_{x_{1}x_{0}|y_{1}}^{(XL)}\left(\varepsilon'_{X}+U-\mu_{XL}\right),\label{eq:qxl}
\end{equation}

\begin{equation}
\dot{Q}_{YR}=-J_{y_{0}y_{1}|x_{0}}^{(YR)}\left(\varepsilon'_{Y}-\mu_{YR}\right)-J_{y_{0}y_{1}|x_{1}}^{(YR)}\left(\varepsilon'_{Y}+U-\mu_{YR}\right),\label{eq:qyr}
\end{equation}
and

\begin{equation}
\dot{Q}_{YL}=J_{y_{1}y_{0}|x_{0}}^{(YL)}\left(\varepsilon'_{Y}-\mu_{YL}\right)+J_{y_{1}y_{0}|x_{1}}^{(YL)}\left(\varepsilon'_{Y}+U-\mu_{YL}\right).\label{eq:qyl}
\end{equation}
The rate of total entropy production can be conveniently related to
the heat currents by summing up Eqs. (\ref{eq:qxr})-(\ref{eq:qyl}),
i.e.,
\begin{equation}
\dot{\sigma}=-\dot{S}_{r}=-\sum_{\nu}\dot{Q}_{\nu}/T\geq0.\label{eq:secl}
\end{equation}

Equations (\ref{eq:gen})-(\ref{eq:el2}) exclusively explain the
entropy flow between the system and the environment. To clarify how
energy and information are exchanged between the two subsystems, we
introduce the rates of partial entropy production associated with
$X$ and $Y$ as \cite{RN11,RN16}

\begin{equation}
\dot{\sigma}_{X}=\sum_{\nu}\sum_{x\geq x^{\prime},y}J_{xx^{\prime}|y}^{(v)}\ln\frac{R_{xx^{\prime}|y}^{(\nu)}p\left(x^{\prime},y\right)}{R_{x^{\prime}x|y}^{(\nu)}p(x,y)},\label{eq:sigmax1}
\end{equation}

\begin{equation}
\dot{\sigma}_{Y}=\sum_{\nu}\sum_{x,y\geq y^{\prime}}J_{yy^{\prime}|x}^{(v)}\ln\frac{R_{yy^{\prime}|x}^{(\nu)}p\left(x,y^{\prime}\right)}{R_{y^{\prime}y|x}^{(\nu)}p(x,y)},\label{eq:sigmax2}
\end{equation}
where the rate of total entropy production is then divided into two
seperate parts followed by

\begin{equation}
\dot{\sigma}=\dot{\sigma}_{X}+\dot{\sigma}_{Y}\text{.}
\end{equation}
These classification can be proved directly from the relation $\sum_{z'}J_{e}=\sum_{\nu,x'}J_{xx'|y}^{(\nu)}+\sum_{\nu,y'}J_{yy'|x}^{(\nu)}$
and the prescribed transition rate {[}Eq. (\ref{eq:tr}){]}. The logarithmic
sum inequality again shows that $\dot{\sigma}_{X}$ and $\dot{\sigma}_{Y}$
are nonnegative individually, i.e.,

\begin{equation}
\dot{\sigma}_{X}\geq0,\qquad\dot{\sigma}_{Y}\geq0,\label{eq:seclaw}
\end{equation}
which is a generalized second law of thermodynamics stronger than
Eq. (\ref{eq:secl}). Equations (\ref{eq:sigmax1}) and (\ref{eq:sigmax2})
dictating the rate of partial entropy production in each subsystem
can be resolved into three components
\begin{equation}
\dot{\sigma}_{X}=\dot{\sigma}_{\mu_{X}}+\dot{\sigma}_{U_{X}}-\dot{I}_{X},
\end{equation}
\begin{equation}
\dot{\sigma}_{Y}=\dot{\sigma}_{\mu_{Y}}+\dot{\sigma}_{U_{Y}}-\dot{I}_{Y},
\end{equation}
where
\begin{align}
\dot{\sigma}_{\mu_{X}} & =\mathcal{\mathscr{\mathcal{F}}}(C_{1}^{X})\left(J\left(C_{1}^{X}\right)+J\left(C_{2}^{X}\right)\right),\nonumber \\
\dot{\sigma}_{\mu_{Y}} & =\mathcal{\mathscr{\mathcal{F}}}(C_{1}^{Y})\left(J\left(C_{1}^{Y}\right)+J\left(C_{2}^{Y}\right)\right),\nonumber \\
\dot{\sigma}_{U_{X}} & =\mathcal{\mathscr{\mathcal{F}}}_{X}\left(C_{1}^{G}\right)J\left(C_{1}^{G}\right),\nonumber \\
\dot{\sigma}_{U_{Y}} & =\mathcal{\mathscr{\mathcal{F}}}_{Y}\left(C_{1}^{G}\right)J\left(C_{1}^{G}\right),\nonumber \\
\dot{I}_{X} & =-J\left(C_{1}^{G}\right)\mathcal{\mathscr{\mathcal{F}}}_{I}\left(C_{1}^{G}\right),\nonumber \\
\dot{I}_{Y} & =J\left(C_{1}^{G}\right)\mathcal{\mathscr{\mathcal{F}}}_{I}\left(C_{1}^{G}\right).\label{eq:partial entropy}
\end{align}
The first term $\dot{\sigma}_{\mu_{X(Y)}}$ indicates that the local
cycle $C_{k}^{X(Y)}$ having a local affinity $\mathcal{\mathscr{\mathcal{F}}}(C_{k}^{X(Y)})$
supports the internal flows within the subsystem. The global cycle
$C_{1}^{G}$ generates a global current $J\left(C_{1}^{G}\right)$
to carry the energy and information from one subsystem to the other.
$\dot{\sigma}_{U_{X}}$ and $\dot{\sigma}_{U_{Y}}$ are responsible
for the direct energy transfer between $X$ and $Y$ due to the Coulomb
coupling. Any energy extracted by the partial affinity $\mathcal{\mathscr{\mathcal{F}}}_{X}\left(C_{1}^{G}\right)$
will be deposited in $Y$\textquoteright s environment by $\mathcal{\mathscr{\mathcal{F}}}_{Y}\left(C_{1}^{G}\right)$.
The information flow $\dot{I}_{X(Y)}$ exclusively occurs on the global
cycle $C_{1}^{G}$ with information affinity $\mp\mathcal{\mathscr{\mathcal{F}}}_{I}\left(C_{1}^{G}\right)$.
When $\mathcal{\mathscr{\mathcal{F}}}_{X}\left(C_{1}^{G}\right)\ll\mathcal{\mathscr{\mathcal{F}}}_{I}\left(C_{1}^{G}\right)$
, the dominant force for driving $X$ is information. On the other
hand, when $\mathcal{\mathscr{\mathcal{F}}}_{X}\left(C_{1}^{G}\right)\gg\mathcal{\mathscr{\mathcal{F}}}_{I}\left(C_{1}^{G}\right)$
, the interaction is mainly powered by energy. The cycle graph analysis
enables a better understanding of the driving mechanisms of the internal
interactions.

\subsection{\textit{Nonequilibrium free energy inequalities of the two subsystems}}

Since $\mathcal{\mathscr{\mathcal{F}}}(C_{1}^{X(Y)})$ takes the chemical
potential difference as the thermodynamic driving force and $J\left(C_{1}^{X(Y)}\right)+J\left(C_{2}^{X(Y)}\right)$
represents the net electronic current along $X(Y)$, the work flux
$W_{X(Y)}$ performed on a subsystem is identitied as

\begin{equation}
\dot{W}_{X(Y)}=T\mathcal{\mathscr{\mathcal{F}}}(C_{1}^{X(Y)})\left[J\left(C_{1}^{X(Y)}\right)+J\left(C_{2}^{X(Y)}\right)\right]=-T\mathcal{\mathscr{\mathcal{F}}}(C_{1}^{X(Y)})J_{X(Y)}.\label{eq:work}
\end{equation}

\noindent According to the first law of thermodynamics, the energy
flow $\dot{E}_{X(Y)}$ into the subsystem reads

\begin{align}
\dot{E}_{X(Y)} & =\dot{Q}_{X(Y)L}+\dot{Q}_{X(Y)R}+\dot{W}_{X(Y)}\nonumber \\
 & =-T\mathcal{\mathscr{\mathcal{F}}}_{X(Y)}\left(C_{1}^{G}\right)J\left(C_{1}^{G}\right),\label{eq:energy flow}
\end{align}
depending on the Coulomb force and the global current. The steady-state
internal energy of the bipartite system remains constant $\left(d_{t}E=\dot{E}_{X}+\dot{E}_{Y}=0\right)$.
Multiplying Eq. (\ref{eq:seclaw}) by $T$ and applying the above
derivatives, one gets

\begin{equation}
T\dot{\sigma}_{X(Y)}=\dot{W}_{X(Y)}-\dot{F}_{X(Y)}\geq0,\label{eq:2SCL}
\end{equation}
where the rate of partial nonequilibrium free energy $\dot{F}_{X(Y)}$
associated with $X$ $\left(Y\right)$ is defined as \cite{RN17,RN18}

\begin{equation}
\dot{F}_{X(Y)}=\dot{E}_{X(Y)}+T\dot{I}_{X(Y)}.
\end{equation}
Equation (\ref{eq:2SCL}) provides an extension of the second law
of thermodynamics. From a practical point of view, it means that the
total amount of work that can be extracted in the nonequilibrium system
is limited by the decrease of free energy. The rate of partial nonequilibrium
free energy can be partitioned into the information-related and energy-related
parts. Its definition has been generally found appropriately, because
$\dot{I}_{X(Y)}$ is equivalent to the rate of the Shannon entropy
of the system $\dot{S}$ due to the interaction with the reservoirs
$XL$ $\left(YL\right)$ and $XR$ $\left(YR\right)$.

\begin{center}
\par\end{center}

\textcolor{red}{}

\section{\textup{RESULTS AND DISCUSSION}}

Without loss of generality, we control over the tunnel rate in the
weak coupling regime. For characterizing the Coulomb blocking effect,
the parameters of the potential barriers are chosen as follows: $\Gamma_{x_{0}}^{\left(XL\right)}=\Gamma\frac{e^{+\delta}}{\cosh\left(\delta\right)}$,
$\Gamma_{x_{1}}^{\left(XL\right)}=\Gamma\frac{e^{-\delta}}{\cosh\left(\delta\right)}$,
$\Gamma_{x_{0}}^{\left(XR\right)}=\Gamma\frac{e^{-\delta}}{\cosh\left(\delta\right)}$,
$\Gamma_{x_{1}}^{\left(XR\right)}=\Gamma\frac{e^{+\delta}}{\cosh\left(\delta\right)}$,
$\Gamma_{y_{0}}^{\left(YL\right)}=\Gamma\frac{e^{+\Delta}}{\cosh\left(\Delta\right)}$,
$\Gamma_{y_{1}}^{\left(YL\right)}=\Gamma\frac{e^{-\Delta}}{\cosh\left(\Delta\right)}$,
$\Gamma_{y_{0}}^{\left(YR\right)}=\Gamma\frac{e^{-\Delta}}{\cosh\left(\Delta\right)}$,
and $\Gamma_{y_{1}}^{\left(YR\right)}=\Gamma\frac{e^{+\Delta}}{\cosh\left(\Delta\right)}$.
\textcolor{black}{In addition, we set $C_{XL}=C_{XR}=C_{YL}=C_{YR}=C_{0}$.}
\begin{center}
\begin{figure}
\includegraphics[scale=0.3]{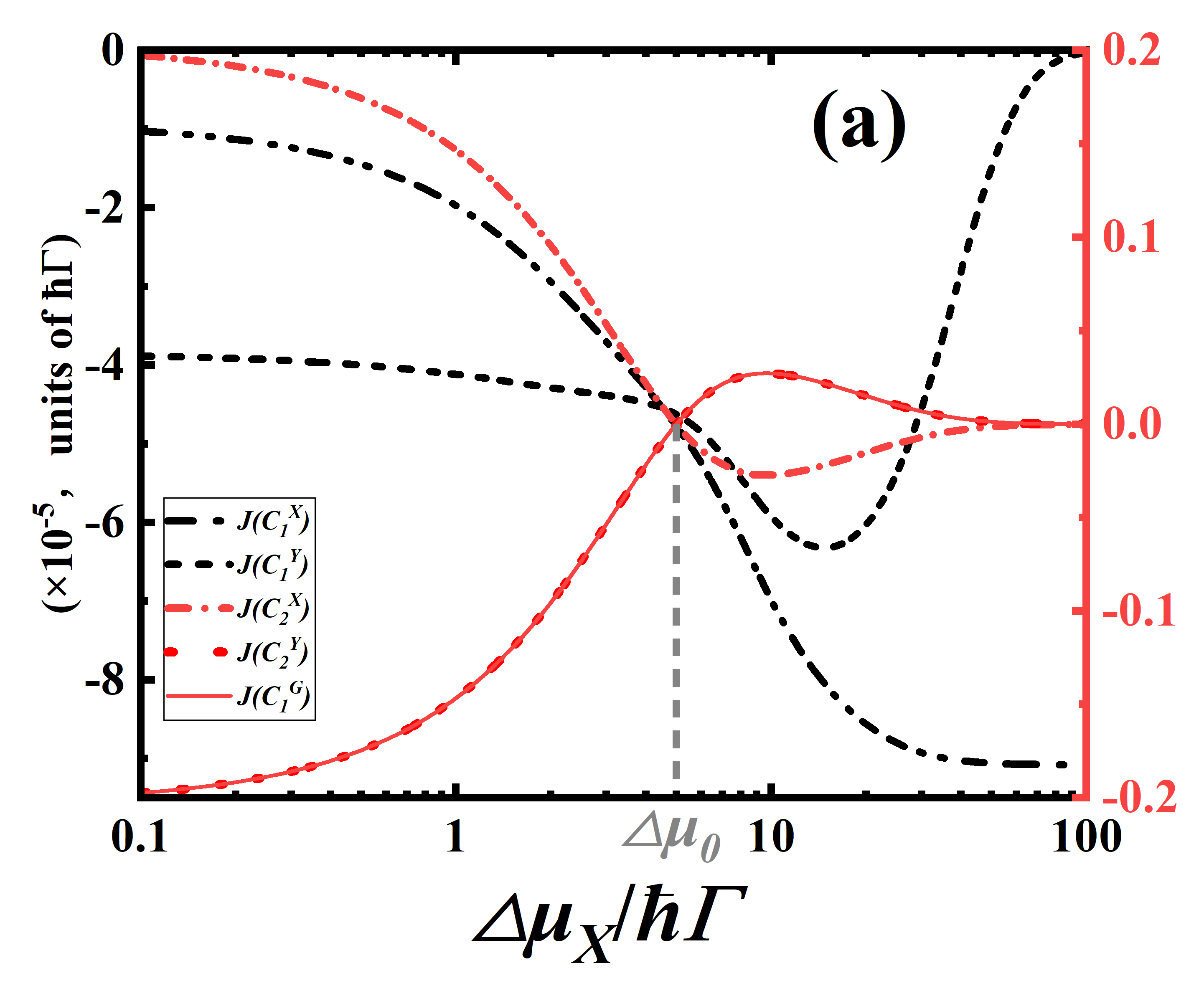}\includegraphics[scale=0.3]{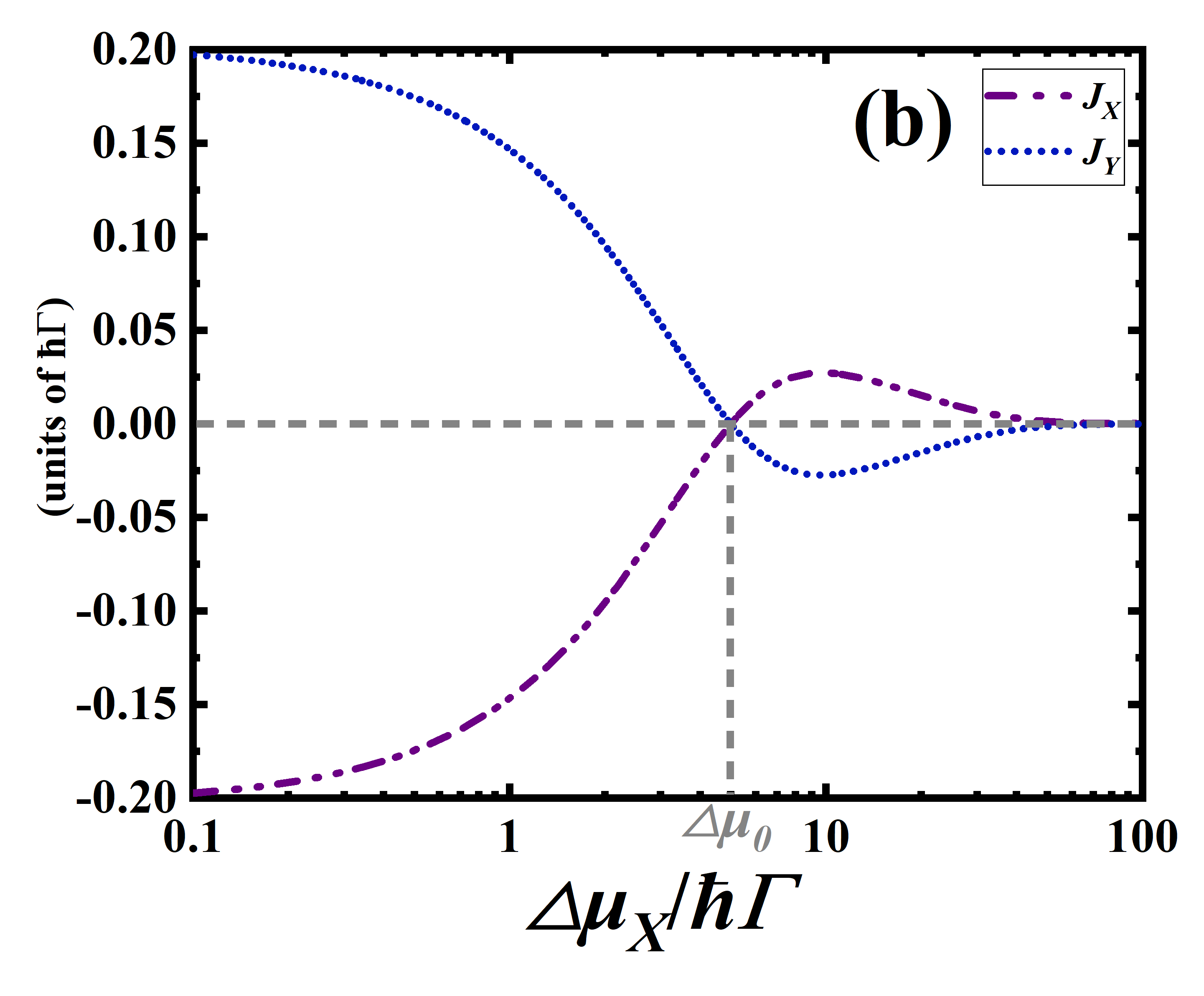}\caption{The curves of (a) the cycle currents and (b) the electronic currents
$J_{X}$ and $J_{Y}$ varying with the chemical potential difference
$\Delta\mu_{X}/\hbar\Gamma$ . The left vertical axis shows the values
for $J(C_{1}^{X})$ and $J(C_{2}^{X})$, while the corresponding scales
of $J(C_{1}^{G})$ , $J(C_{1}^{Y})$, and $J(C_{2}^{Y})$ are on the
right vertical axis. The parameters $\varepsilon_{X}=0$, $\varepsilon_{Y}=0$,
$\Delta\mu_{Y}=5.0\hbar\Gamma$, $\mu_{XL}-\mu_{YL}=1.0\hbar\Gamma$,
$C=0.5q^{2}/\hbar\Gamma$, $C_{0}=1.0q^{2}/\hbar\Gamma$, $k_{B}T=1.0\hbar\Gamma$,
and $\Delta=\delta=5.0$. These values are used unless otherwise mentioned
specifically in the following discussion.\label{fig:cycle current}}
\end{figure}
\par\end{center}

\textcolor{black}{Figure \ref{fig:cycle current} (a) shows the currents
of the fundamental cycles selected in Eq. (\ref{eq:cc}). When the
dot $Y$ $(X)$ is empty, the local cycle current $J(C_{1}^{X})<0$
(black dash-double-dotted line) {[}$J(C_{1}^{Y})<0$ (black dashed
line){]}, meaning that electrons are prone to transfer from the higher
chemical potential to the lower chemical potential. When $Y$ is occupied,
the local cycle current $J(C_{2}^{X})$ (red dash-dotted line) through
$X$ does not move in a fixed direction, but is embodied in two situations.
The first case is characterized by $J(C_{2}^{X})>0$ in the small-$\Delta\mu_{X}$
regime ($\Delta\mu_{X}<\Delta\mu_{0}$), where the electrons transfer
from reservoir $XL$ to reservoir $XR$ against the thermodynamic
force $\mathcal{\mathscr{\mathcal{-F}}}(C_{2}^{X})$ due to the Coulomb
coupling. The second case is characterized by $J(C_{2}^{X})<0$ in
the large-$\Delta\mu_{X}$ regime ($\Delta\mu_{X}>\Delta\mu_{0}$)),
because the thermodynamic force $\mathcal{-\mathscr{\mathcal{F}}}(C_{2}^{X})$
dominates the electron transport again. When $X$ is occupied, the
local cycle current $J(C_{2}^{Y})$ (red dashed line) }changes\textcolor{black}{{}
from negative to positive as $\Delta\mu_{X}$ increases. When $\Delta\mu_{X}$
is large enough, the dot $X$ may provide enough energy or information
for driving the electron flow in the}\textcolor{red}{{} }direction\textcolor{black}{{}
opposite to the chemical potential gradient $\mathcal{\mathscr{\mathcal{-F}}}(C_{2}^{Y})$.
The graph theory offers an effective way to unearthing the fundamental
path, where the current through one dot could drag the current through
the other dot. }

\textcolor{black}{In the regime of small $\Delta\mu_{X}$, the direction
of the global cycle current is counterclockwise $J(C_{1}^{G})<0$.
According to Eq. (\ref{eq:energy flow}), multiplying $J(C_{1}^{G})$
by the partial affinities $\mathcal{-\mathscr{\mathcal{F}}}_{X}\left(C_{1}^{G}\right)$
and $\mathcal{-\mathscr{\mathcal{F}}}_{Y}\left(C_{1}^{G}\right)$,
respectively, we know that the energy flow is fed from $X$ }to $Y$\textcolor{black}{{}
(}grey dash-double-dotted lines\textcolor{black}{{} Fig. 4). In the
latter, the direction of the global cycle current becomes clockwise
as $\Delta\mu_{X}$ increases, which lies the regime suitable for
the energy flowing from $Y$ }to $X$\textcolor{black}{. By applying
Eqs. (\ref{eq:cx}) and (\ref{eq:cy}), Fig. \ref{fig:cycle current}
(b) presents the net electronic currents from $XR$} to\textcolor{black}{{}
$XL$ $\left(J_{X}\right)$ and from $YR$} to\textcolor{black}{{} $YL$
$\left(J_{Y}\right)$. Owing to the existence of electron-electron
interaction, the local cycle current $J(C_{2}^{X})$ $\left[J(C_{2}^{Y})\right]$
makes a major contribution to the currents $J_{X}$ $\left(J_{Y}\right)$.
For $\Delta\mu_{X}<\Delta\mu_{0}$, the net electronic current $J_{X}$
flows from the reservoir $XL$ with the lower chemical potential to
the reservoir $XR$ with the higher chemical potential. The work flux
$\dot{W}_{X}<0$ (black solid line in Fig. 4), so that the dot $X$
produces power output. As $\Delta\mu_{X}$ gets larger, the situation
is exactly the opposite. The net electronic current $J_{Y}$ flows
against the bias due to the thermodynamic force and the dot $Y$ starts
to generate positive power output.}
\begin{center}
\begin{figure}
\includegraphics[scale=0.5]{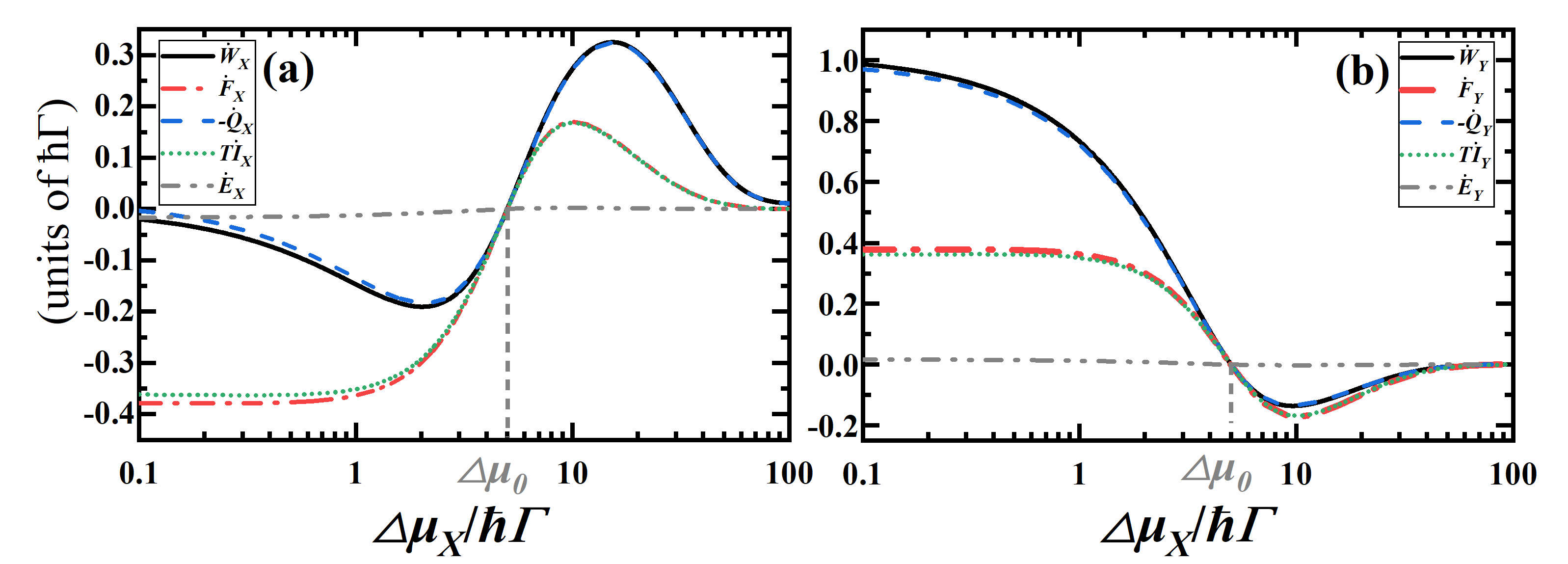}\caption{The work fluxes (black solid lines), rates of partial nonequilibrium
free energy (red dash-dotted lines), heat currents (blue dashed lines),
information flows (green dotted lines), and energy flows (grey dash-double-dotted
lines) of the subsystems (a) $X$ and (b) $Y$ varying with the chemical
potential difference $\Delta\mu_{X}/\hbar\Gamma$ at steady state.
\label{fig:The-work-fluxes}}
\end{figure}
\par\end{center}

\textcolor{black}{The currents of the fundamental cycles are more
important to investigate the thermodynamics of }the rates of the partial
nonequilibrium free energy (red dash-dotted lines) and information
flows (green dotted lines)\textcolor{black}{.} The nonequilibrium
Clausius and free energy inequalities {[}Equations (\ref{eq:seclaw})
and (\ref{eq:2SCL}){]} serve the subsystems with the local second
law of thermodynamics, which are demonstrated in Fig. \ref{fig:The-work-fluxes}.
They behave equivalently and complement each other for a unify treatment
of temperature. \textcolor{black}{For $\Delta\mu_{X}<\Delta\mu_{0}$,
the total heat current into $X$, $\dot{Q}_{X}=\dot{Q}_{XR}+\dot{Q}_{XL}>0$
. To achieve this effect, a noticeable information flowing} from $Y$
to $X$ \textcolor{black}{guarantees that $T\dot{\sigma}_{X}=-\dot{Q}_{X}-T\dot{I}_{X}>0$.
The information flow benefits from the current of the global cycle
$J\left(C_{1}^{G}\right)$ driven by the information affinity $\mathcal{\mathscr{\mathcal{F}}}_{I}\left(C_{1}^{G}\right)$
. At the same time, the dot $X$ performs work $\dot{W}_{X}<0$, which
is enabled by the negative rate of partial nonequilibrium free energy
$0>\dot{W}_{X}>\dot{F}_{X}$. Note that these effects are compensated
by the dissipation of work into heat in the dot $Y$. For }$\Delta\mu_{X}\text{>}\Delta\mu_{0}$,
the role\textcolor{black}{s of $X$ and $Y$ are totally rev}ersed.
It should be emphasized that under most conditions the power output
is generated due to the feedback-induced information flow and not
due to energy flow, because $\dot{E}_{X(Y)}\approx0$ and $\dot{F}_{X(Y)}\approx T\dot{I}_{X(Y)}$.
As a result, the nonequilibrium double quantum-dot system works as
a quantum autonomous Maxwell demon.

\section{\textup{CONCLUSIONS}}

In summary, the present work reveals the local second law of thermodynamics
of the Coulomb-coupled double quantum dots. The graph theory has proved
that the entropy production and nonequilibrium free energy are close
related to the information flow and affinities. The information flow
between the subsystems acts as a driving force for the global cycle
to pump electrons against the bias voltage. The proposed model offers
possible schemes to design nanoelectronic devices through the control
of cycle fluxes.

\begin{acknowledgments}
This work has been supported by the National Natural Science Foundation (Grant No. 11805159), the Fundamental Research Fund for the Central Universities (No. 20720180011), and the Natural Science Foundation of Fujian Province of China (No. 2019J05003).
\end{acknowledgments}

\end{document}